\documentclass{PoS}
\renewcommand{\Re}{\mbox{Re }}
\title{From realistic 2HDM-II CPV benchmarks\\
to the $H^\pm \to\tau \nu$ decay at the LHC}

\ShortTitle{From realistic 2HDM-II CPV benchmarks\dots}

\author{Lorenzo Basso$^{ab}$, Per Osland$^c$ and \speaker{Giovanni Marco Pruna}\thanks{This work has been supported by the European Community's Seventh Framework Programme (FP7/2007-2013) under grant agreement n.~290605 (COFUND: PSI-FELLOW).}$^{\ d}$ \\ \ \\
\llap{$^a$} Universit\'e de Strasbourg, IPHC,\\
            23 rue du Loess 67037 Strasbourg, France \\ \ \\
\llap{$^b$} CNRS, UMR7178,\\
            67037 Strasbourg, France \\ \ \\
\llap{$^c$} Department of Physics and Technology, University of Bergen, \\
        Postboks 7803, N-5020  Bergen, Norway\\ \ \\
\llap{$^d$}
        Paul Scherrer Institut, \\
        CH-5232 Villigen PSI, Switzerland\\ \ \\
E-mails: \email{Lorenzo.Basso@iphc.cnrs.fr}\\
\phantom{E-mails: }\email{Per.Osland@ift.uib.no}\\
\phantom{E-mails: }\email{Giovanni-Marco.Pruna@psi.ch}}

\abstract{Phenomenological studies of a CP-violating two-Higgs-doublet Model with type-II Yukawa couplings are presented. In the light of recent LHC data, an update on the viable parameter space that survives both the experimental and theoretical constraints is provided. In addition, the scope of the LHC in exploring this model through the discovery of a charged Higgs boson that decays in the tauonic mode is analysed. For this, various production channels were investigated, with emphasis on the boson-associated channel $gg\to H_i\to H^\pm W^\mp$ and the fermion-associated channels $gb\to H^\pm t$ and $gg\to H^\pm bt$. For the latter, a signal-over-background analysis is performed.}

\FullConference{Prospects for Charged Higgs Discovery at Colliders - CHARGED 2014,\\
		16-18 September 2014\\
		Uppsala University, Sweden}

\begin{document}

\section{Introduction}
\noindent
After the Higgs boson discovery \cite{Aad:2012tfa,Chatrchyan:2012ufa}, tests concerning the minimality of the Higgs sector of the Standard Model (SM) assume a leading role in the present and future experimental programmes at colliders. Among the various phenomenological investigations, searches for new scalar resonances are generally considered a top priority, as they would directly probe the existence and the features of non-minimal Higgs contents, possibly establishing connection with physics beyond the SM of particles, \emph{e.g.} a Minimal Supersymmetric Standard Model (MSSM) particle content.

Here, the popular option of an extended scalar sector containing two Higgs doublets is considered, \emph{i.e.,} the two-Higgs-doublet Model (2HDM) \cite{Gunion:1989we} is studied. If such a model is realised at the TeV scale, it is well-known that one should expect the existence of exotic resonances such as charged scalars to be probed at the Large Hadron Collider (LHC). In its general formulation \cite{Accomando:2006ga}, a 2HDM allows for a mixing among CP-odd and CP-even components of the scalar particles. Furthermore, many realisations of the Yukawa sector are equally available. In particular, the production of a charged Higgs of a 2HDM with mixed CP contents and a type-II Yukawa sector (motivated by the MSSM-like structure), referred to as ``CPV 2HDM type-II'' \cite{Khater:2003wq}, is investigated in the context of the LHC run 2. Among the possible decay modes, the focus is set on the tauonic decay, due to the fact that an interaction of the charged Higgs with the third generation leptons would represent a striking probe of the Yukawa structure of such a model.

\section{The Model}
\noindent
In this Section, the parametrisation of the CPV 2HDM type-II is described. As previously intimated, the Higgs sector is defined by the presence of two Higgs doublets, with one ($\Phi_2$) coupled to the $u$-type quarks, and the other one ($\Phi_1$) to the $d$-type quarks and charged leptons, in analogy with the MSSM. The scalar potential is described as follows:
\begin{eqnarray}
\label{Eq:pot_7}
V&=&\frac{\lambda_1}{2}(\Phi_1^\dagger\Phi_1)^2
+\frac{\lambda_2}{2}(\Phi_2^\dagger\Phi_2)^2
+\lambda_3(\Phi_1^\dagger\Phi_1) (\Phi_2^\dagger\Phi_2) \nonumber \\
&+&\lambda_4(\Phi_1^\dagger\Phi_2) (\Phi_2^\dagger\Phi_1)
+\frac{1}{2}\left[\lambda_5(\Phi_1^\dagger\Phi_2)^2+{\rm h.c.}\right] \nonumber \\
&-&\frac{1}{2}\left\{m_{11}^2(\Phi_1^\dagger\Phi_1)
\!+\!\left[m_{12}^2 (\Phi_1^\dagger\Phi_2)\!+\!{\rm h.c.}\right]
\!+\!m_{22}^2(\Phi_2^\dagger\Phi_2)\right\}, 
\end{eqnarray}
where $\lambda_5$ and $m_{12}^2$ are complex parameters which trigger the mixing between CP-odd and CP-even components \cite{Ginzburg:2002wt}. Following the parametrisation of \cite{ElKaffas:2006nt,ElKaffas:2007rq}, the starting number of degrees of freedom is $8$: $3$ of them represent the would-be Goldstone components of the massive gauge bosons, whilst the remaining $5$ are indeed the physical scalar objects of the theory: $H_1$, $H_2$, $H_3$ and $H^\pm$. The whole parameter space is described by $8$ parameters: $\tan\beta$ (ratio of vacuum expectation values (VEVs) of the doublets), $\sin{\alpha_i}$ (mixing parameters, with $i=1,3$), $M_1$ (mass of $H_1$), $M_2$ (mass of $H_2$), $M_{H^\pm}$ (mass of $H^\pm$), $\mu$ (combination of Lagrangian mass parameters and VEVs: $\mu^2=\Re m_{12}^2/(2\cos\beta\sin\beta)$). (The mass of the third neutral Higgs boson is determined by these parameters \cite{Khater:2003wq}.)
As a matter of fact, CP violation potentially modifies the structure of the Higgs interactions, hence Yukawa couplings could also be affected. 

\section{Parameter space}
\noindent
In continuity with \cite{Basso:2012st,Basso:2013wna}, updates on the surviving multi-dimensional type-II 2HDM parameter space are given. The following theoretical (\textbf{T}) and experimental (\textbf{E}) constraints were considered:
\begin{itemize}
\item[\textbf{T}:] positivity, tree-level perturbative unitarity, perturbativity, global minimum of the scalar potential.
\item[\textbf{E}:] $B\to X_s \gamma$, $B_u\to \tau \nu_\tau$, $B\to D\tau \nu_\tau$, $D_s\to \tau \nu_\tau$, $B_{d,s}\to \mu^+\mu^-$, $B^0-\overline{B}^0$ (the SM predictions for the flavour observables are obtained using SuperIso v3.2 \cite{Mahmoudi:2008tp}), $R_b$, $pp \to H_jX$ (LHC constraints), $T$ and $S$ (EW precision tests), Electron Electric Dipole Moment (EDM).
\end{itemize}
Although most of them remain unchanged with respect to the previous analysis, LHC constraints received various updates. As a result of the frequent releases of experimental results, they are treated with great caution. The signal strengths $\mu_{\gamma\gamma}$, $\mu_{ZZ}$ and $\mu_{\tau\tau}$ are evaluated, and parameter points violating any one of these by more than $3\,\sigma$ \cite{Heinemeyer:2013tqa}\footnote{See the ATLAS and CMS TWiki pages for updates.} are systematically excluded. The couplings of $H_2$ and $H_3$ to $WW$ are evaluated, and only parameter points corresponding to non-discovery \cite{Chatrchyan:2013yoa,Aad:2013dza} of such heavier states are kept. Even if a substantial prudence is taken into account, LHC data are manifestly pointing towards an aligned scenario where the $H_1$ is assuming more and more SM-like Higgs features. 

Having implemented these constraints, and using ``physical'' input in terms of mass parameters and mixing angles as described in \cite{Khater:2003wq}, selected discrete values of $\tan\beta$, $M_2$, $M_{H^\pm}$, and $\mu$, each with a scan over 5 million trial sets of mixing angles, $\{\alpha_1,\alpha_2,\alpha_3\}$ were sampled. In the next section, these points are analysed with emphasis on the phenomenology of the charged Higgs in both hadronic production and tauonic decay.

\section{Production plus tauonic decay of a charged Higgs and significance at the LHC}
\noindent
In this Section, the phenomenology of the charged Higgs decaying in the $\tau+\nu_\tau$ mode is ana\-lysed in the framework of the LHC run 2. A certain level of automation was required, hence the following available tools were exploited: the Lagrangian of the model was implemented in {\tt LanHEP v3.1.9} \cite{Semenov:2010qt}\footnote{The Higgs sector of the model was implemented in LanHEP according to the description in \cite{Mader:2012pm}, while the Yukawa sector was borrowed from \cite{Basso:2012st}.}; the calculation of cross sections and branching fractions as well as the generation of events for the signal was done in {\tt CalcHEP v3.4.6}~\cite{Belyaev:2012qa} with the CTEQ6L PDF set~\cite{Pumplin:2002vw}; the generation of the background events was performed with {\tt MadGraph5\underline{ }aMC@NLO v2.1.2}~\cite{Alwall:2014hca} employing the CTEQ6L1 PDF set; the event analysis was done with the package {\tt Madanalysis~5 v1.1.11}~\cite{Conte:2012fm}.

The main experimental scenario considered is a hadronic collider with $E=14$ TeV and $L=100$ fb$^{-1}$, according to the ``LHC run 2'' prototype. At the partonic level, the relevant charged Higgs production channels in hadronic collisions are: $qq'\to W^\pm \to H^\pm H_i$, $gg\to H_i\to H^\pm W^\mp$ (boson-associated productions), $g\bar{b}\to H^+ \bar{t}$ (and charge conjugated) and $gg\to H^+ b\bar{t}$ (and charge conjugated) (fermion-associated productions). The off-shell $W$-mediated production is subdominant in this context, thus the following treatment will focus on the three other channels.

\begin{figure}[!ht]
\center
\includegraphics[width=.45\textwidth]{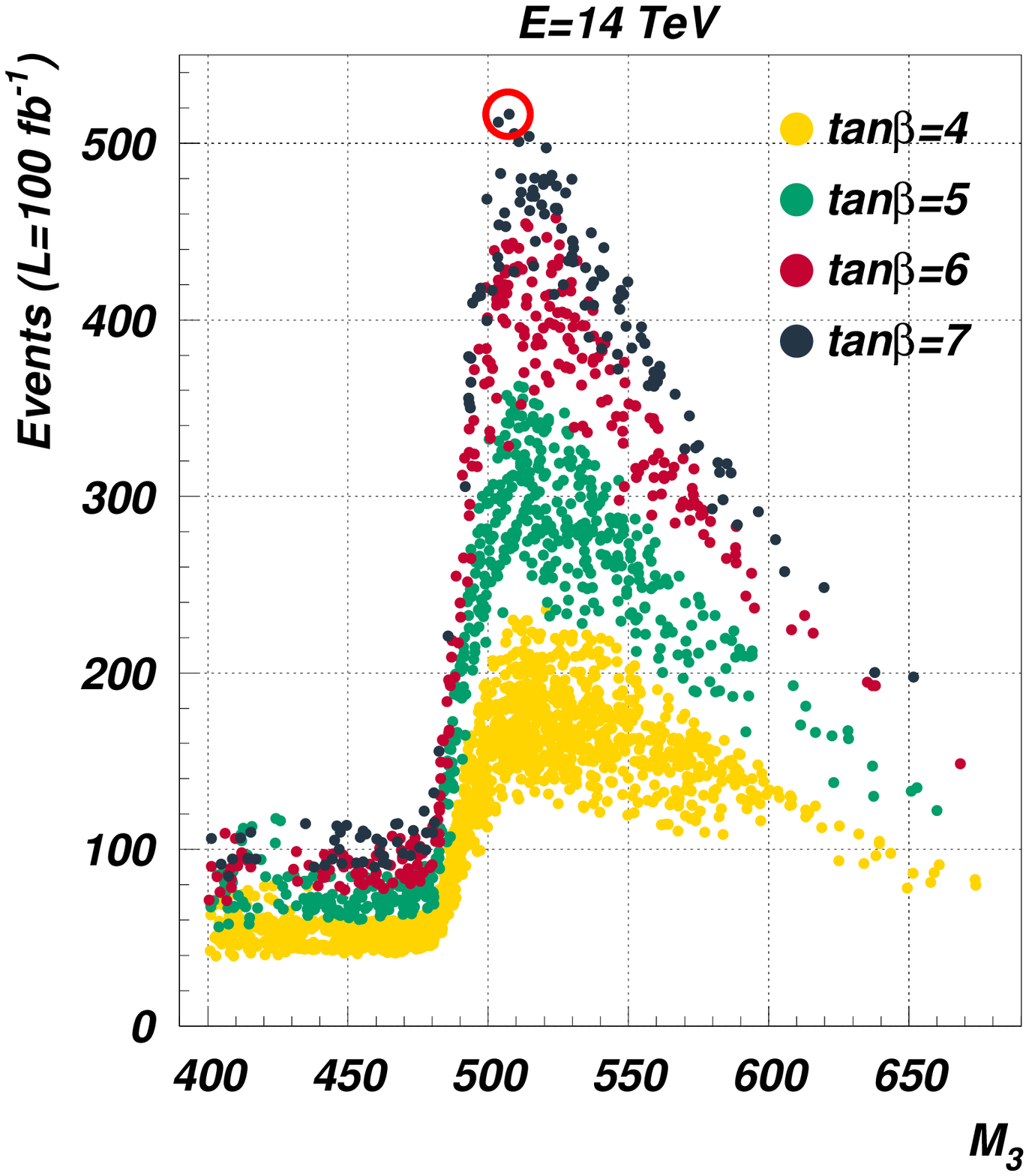}
\includegraphics[width=.45\textwidth]{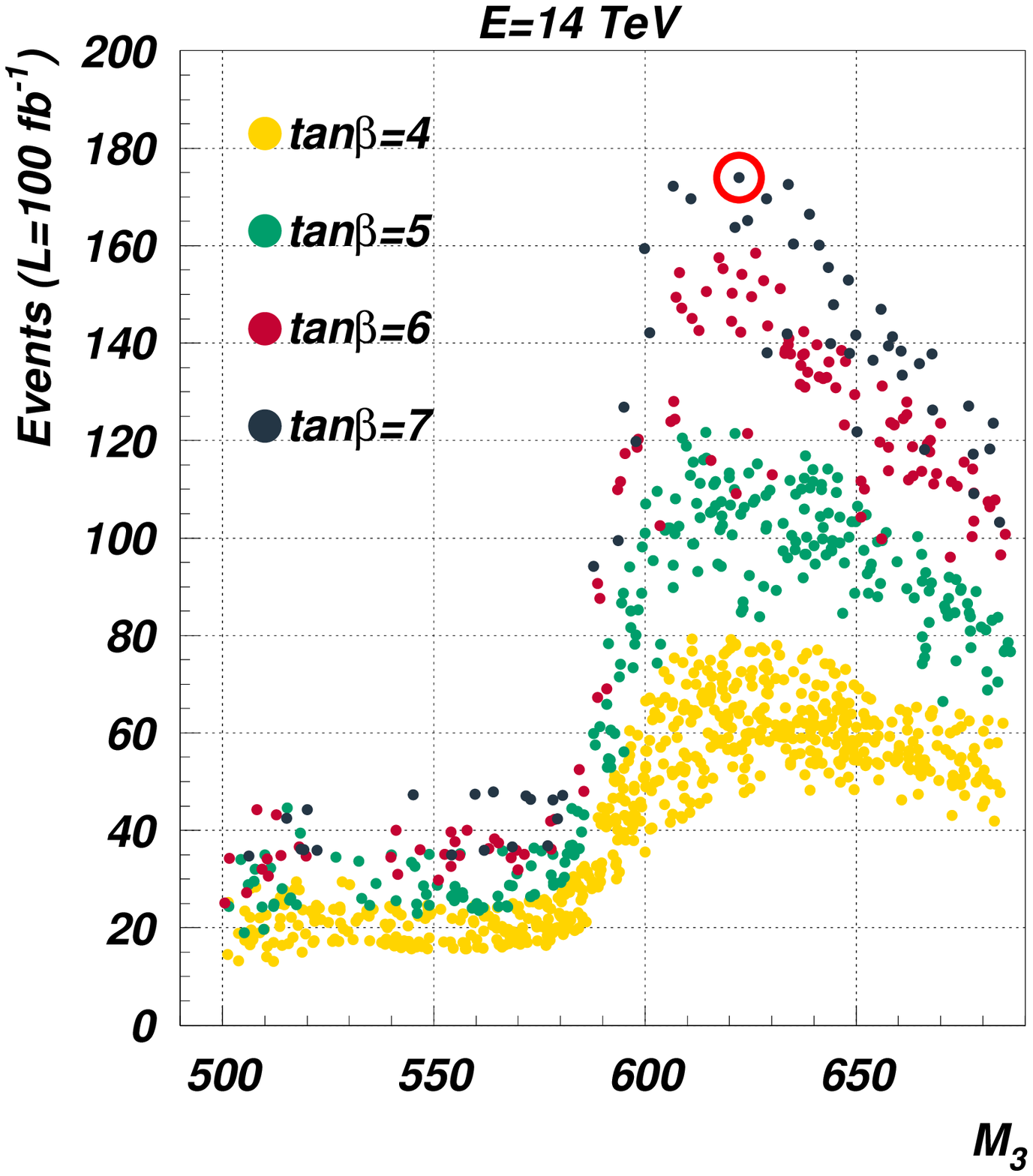}
\vskip -0.5cm
\caption{Number of events produced via $gg\to H_i\to H^\pm W^\mp\to \tau\nu W$ vs $M_{3}$ at a collision energy of $E=14$~TeV and a luminosity of $L=100$~fb$^{-1}$ for various benchmarks with $M_{H^\pm}=400$~GeV (left plot) and $M_{H^\pm}=500$~GeV (right plot). Red circles indicate the best points.}
\label{fig1}
\end{figure}

\begin{figure}[!ht]
\center
\includegraphics[width=.45\textwidth]{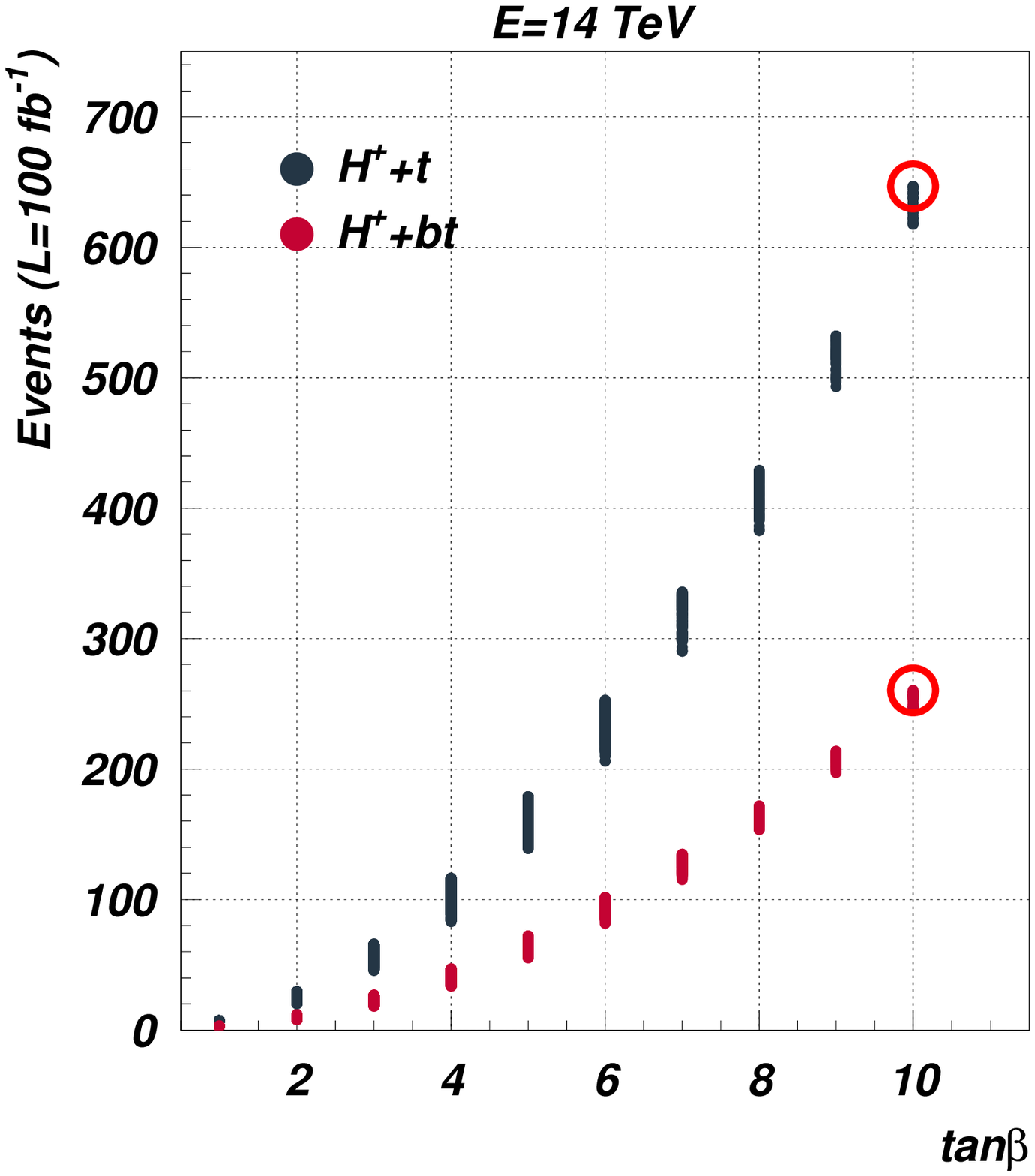}
\includegraphics[width=.45\textwidth]{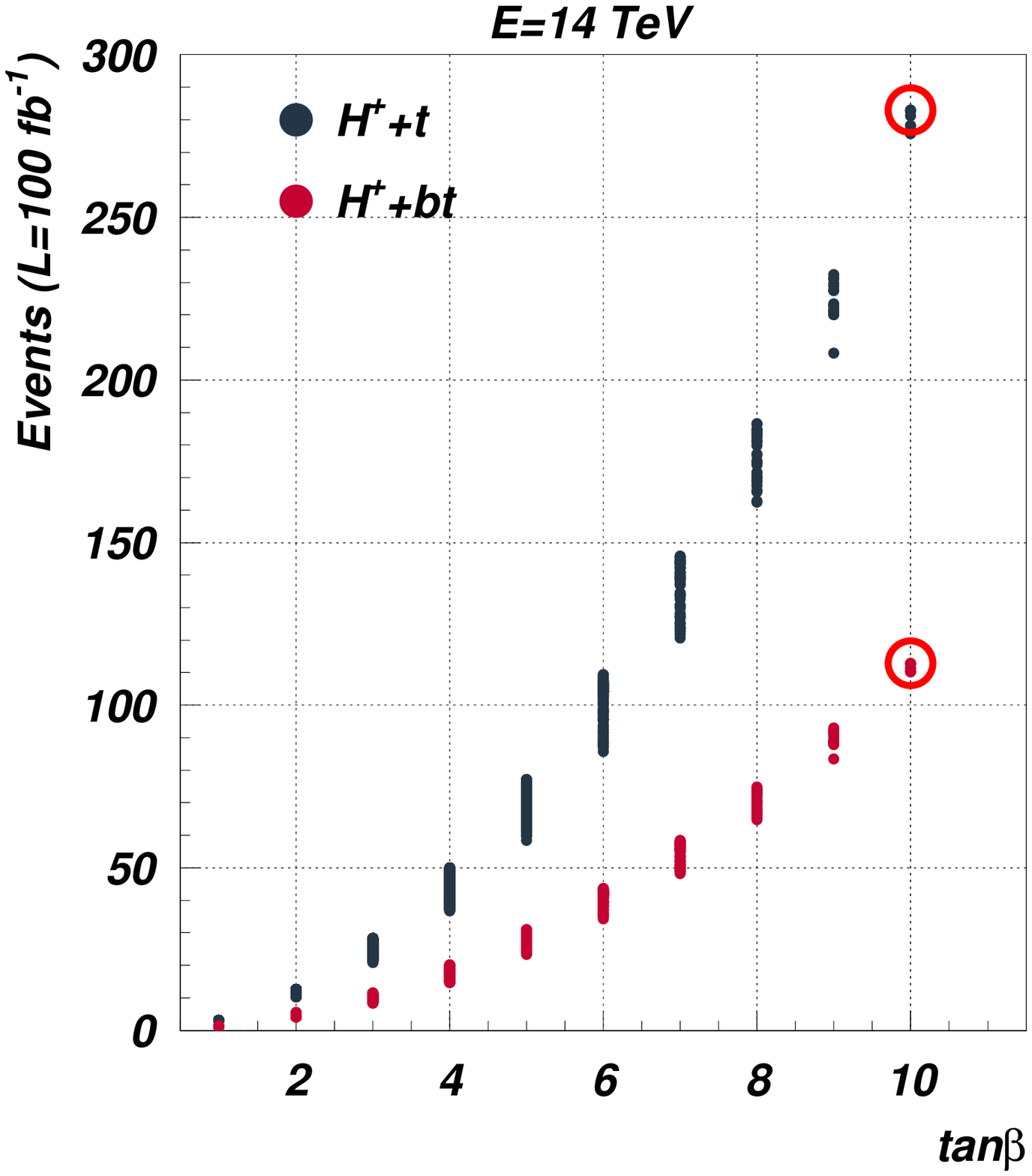}
\vskip -0.5cm
\caption{Number of events produced via $pp\to H^\pm t(b)\to \tau\nu t(b)$ vs $\tan\beta$ at a collision energy of $E=14$ and a luminosity of $L=100$~fb$^{-1}$ for various benchmarks with $M_{H^\pm}=400$~GeV (left plot) and $M_{H^\pm}=500$~GeV (right plot). Red circles indicate the best points.}
\label{fig2}
\end{figure}

In general, the production rate is maximised when the mass of the charged Higgs $M_{H^\pm}$ is as light as possible. However, the limits from $B\to X_s \gamma$ \cite{Hermann:2012fc} push the $M_{H^\pm}$ of this model to be greater than $380$ GeV. This information leads to the choice of considering a charged Higgs mass in the range  $400<M_{H^\pm}/{\rm [GeV]}<500$. Then, configurations that reproduce a certain alignment were preferred, \emph{e.g.} choices like $M_2\simeq \mu\simeq M_{H^\pm}$, $\tan{\beta}\gg 1$, $\sin{\alpha_1}\sim 1$ and $\sin{\alpha_2}\sim 0$. In this framework, the analysis of the main three production rates plus tauonic decays was performed. The ``boson-associated'' and the ``fermion-associated'' productions are separately presented in Figures~\ref{fig1}~and~\ref{fig2}, respectively. 

In Figure~\ref{fig1}, the expected number of events for the boson-associated production plus tauonic decay is plotted against $M_3$ in the framework of LHC run 2 for various choices of $\tan{\beta}$. The maximum reach occurs when the latter is around $\tan{\beta}=7$ and the production becomes resonant with respect to $H_3$, \emph{i.e.} $gg\to H_3\to H^\pm W^\mp$. In Figure~\ref{fig2}, the expected number of events for the fermion-associated production plus tauonic decay is plotted against $\tan{\beta}$, clearly showing that the cross sections for such processes are almost independent from other parameters.
\begin{table}[!ht]
\begin{center}
\begin{tabular}{|c|c|c|c|c|c|c||c|}
\hline
\ & $\alpha_1/\pi$ & $\alpha_2/\pi$ & $\alpha_3/\pi$ & $\tan\beta$ & $M_2$ & $M_{H^\pm}$ & Production\\
\hline
 $P_1$ & $1.42953$ & $-0.01299$ & $0.11118$ & $7$ & $400$ & $400$ & Bosonic\\
 $P_2$ & $1.43129$ & $-0.01909$ & $0.18063$  & $7$ & $500$ & $500$ & Bosonic\\
\hline
 $P_3$ & $1.48311$ & $-0.01026$ & $0.10666$ & $10$ & $400$ & $400$ & Fermionic\\
 $P_4$ & $1.46942$ & $-0.00928$ & $0.13918$ & $10$ & $500$ & $500$ & Fermionic\\
\hline
\end{tabular}
\end{center}
\vskip -0.5cm
\caption{Selected benchmark points. Masses are in GeV. $\mu$ is chosen $\sim M_2$.\label{points}}
\end{table}

The cross sections for the best points read as follows: for the boson-associated production, if $M_{H^\pm}=400$ ($500$) GeV, one has $\sigma=5.16$ ($1.74$)~fb; for the fermion-associated production, if $M_{H^\pm}=400$ ($500$) GeV, one has $\sigma=6.45$ ($2.83$)~fb. In Table~\ref{points}, such points are identified with a label and the related values of the parameters are collected.

Next, the significance of the fermion-associated signal is studied.
The background mainly consists of three final states: $W+3j$, $t+W+j$, and $t+W$, listed from the most to the least relevant. For the case in which the final state includes more than $1$ jet, the following cuts were applied to ensure the convergence of the integration: $p_T^j > 20$ GeV, $|\eta _j| < 5$ $\forall j$, $\Delta R(jj)>0.1$ and $10 < M_{jj}/{\rm [GeV]} < 180$. The inclusive cross-sections are $\sigma_{\tau\nu 3j}=3110$ pb, $\sigma_{t\tau\nu j}=57$ pb and $\sigma_{t\tau\nu}=4.52$ pb. For the reconstruction of the signal, the final state was required to consist of $1$ $\tau$, $N\geq 3$ jets of which exactly one is $b$-tagged, plus the following cuts were implemented: $p_T^j,p_T^\tau > 40$ GeV, $\left|\eta _j\right| < 2.4$ $\forall j$, $\left|\eta _\tau\right| < 2.3$, $\Delta R(jj)>0.5$, $\Delta R(\tau j)>0.3$ $\forall j$, $b(c)$-tagging efficiency of $70\%$ ($20\%$) \cite{Chatrchyan:2012jua}, mistagging rate for light jets of around $1\%$, flat $\tau$-tagging efficiency 25\% \cite{Chatrchyan:2012zz}. Furthermore, $\mbox{MET} > 100 \mbox{ GeV}$ was chosen to cut $W\to\tau\nu$, $\left| M_{jj} - M_W\right| < 30 \mbox{ GeV}$ was imposed to select jets stemming from a $W$, $\left|M_{jjj}-M_t\right| < 30 \mbox{ GeV}$ was set to reconstruct the intermediate $t$-quark. For the $M_{H^\pm}=400$ ($500$) GeV case, the peak was selected with a cut of $350$ ($450$)$ < M_T(\tau)/{\rm [GeV]} < 420$ ($520$).

Despite the good efficiency of the strategic cuts, summing up the $tbH^\pm$ and the $tH^\pm$ modes one obtains the following number of events for $M_{H^\pm}=400$ ($500$) GeV: $S=9.3$ ($3.9$), $B_{t\tau\nu}=0.6$ ($0.1$), $B_{t\tau\nu j}=1.4$ ($0.2$), $B_{\tau\nu 3j}=2.5$ ($1.2$). This leads to a signal-over-background significance of $\Sigma=S/\sqrt{S+B}=2.5$ ($1.7$).

\section{Conclusion}
\noindent
The production and tauonic decay of a CPV 2HDM type-II charged Higgs at the LHC run 2 was presented. The main production modes were analysed, and the significance of the fermion-associated production was determined for chosen benchmark points, turning out to be below the discovery threshold. For this, the studied production mode calls for either a luminosity or energy upgrade \cite{Bruning:2002yh}.
The boson-associated case needs further study.


\begin{thebibliography}{99}

\bibitem{Aad:2012tfa}
  G.~Aad {\it et al.}  [ATLAS Collaboration],
  \emph{Phys.\ Lett.\ B} {\bf 716} (2012) 1
  [{\tt arXiv:1207.7214}].

\bibitem{Chatrchyan:2012ufa}
  S.~Chatrchyan {\it et al.}  [CMS Collaboration],
  \emph{Phys.\ Lett.\ B} {\bf 716} (2012) 30
  [{\tt arXiv:1207.7235}].

\bibitem{Gunion:1989we}
  J.~F.~Gunion, H.~E.~Haber, G.~L.~Kane and S.~Dawson,
  \emph{Front.\ Phys.\ }  {\bf 80} (2000) 1.

\bibitem{Accomando:2006ga}
  E.~Accomando, A.~G.~Akeroyd, E.~Akhmetzyanova, J.~Albert, A.~Alves, N.~Amapane, M.~Aoki and G.~Azuelos {\it et al.},
  [{\tt hep-ph/0608079}].

\bibitem{Khater:2003wq}
  W.~Khater and P.~Osland,
  \emph{Nucl.\ Phys.\ B} {\bf 661} (2003) 209
  [{\tt hep-ph/0302004}].

\bibitem{Ginzburg:2002wt}
  I.~F.~Ginzburg, M.~Krawczyk and P.~Osland,
  In *Seogwipo 2002, Linear colliders* 90-94
  [{\tt hep-ph/0211371}].

\bibitem{ElKaffas:2006nt}
  A.~W.~El Kaffas, W.~Khater, O.~M.~Ogreid and P.~Osland,
  \emph{Nucl.\ Phys.\ B} {\bf 775} (2007) 45
  [{\tt hep-ph/0605142}].

\bibitem{ElKaffas:2007rq}
  A.~W.~El Kaffas, P.~Osland and O.~M.~Ogreid,
  \emph{Nonlin.\ Phenom.\ Complex Syst.\ } {\bf 10} (2007) 347
  [{\tt hep-ph/0702097}].

\bibitem{Basso:2012st}
  L.~Basso, A.~Lipniacka, F.~Mahmoudi, S.~Moretti, P.~Osland, G.~M.~Pruna and M.~Purmohammadi,
  \emph{JHEP} {\bf 1211} (2012) 011
  [{\tt arXiv:1205.6569}].

\bibitem{Basso:2013wna}
  L.~Basso, A.~Lipniacka, F.~Mahmoudi, S.~Moretti, P.~Osland, G.~M.~Pruna and M.~Purmohammadi,
  \emph{PoS Corfu} {\bf 2012} (2013) 029
  [{\tt arXiv:1305.3219}],
  \pos{PoS(Corfu2012)029}.

\bibitem{Mahmoudi:2008tp}
  F.~Mahmoudi,
  \emph{Comput.\ Phys.\ Commun.\ } {\bf 180} (2009) 1579
  [{\tt arXiv:0808.3144}].

\bibitem{Heinemeyer:2013tqa}
  S.~Heinemeyer {\it et al.}  [LHC Higgs Cross Section Working Group Collaboration],
  [{\tt arXiv:1307.1347}].

\bibitem{Chatrchyan:2013yoa}
  S.~Chatrchyan {\it et al.}  [CMS Collaboration],
  \emph{Eur.\ Phys.\ J.\ C} {\bf 73} (2013) 2469
  [{\tt arXiv:1304.0213}].

\bibitem{Aad:2013dza}
  G.~Aad {\it et al.}  [ATLAS Collaboration],
  \emph{Phys.\ Rev.\ D} {\bf 89} (2014) 3,  032002
  [{\tt arXiv:1312.1956}].

\bibitem{Mader:2012pm}
  W.~Mader, J.~h.~Park, G.~M.~Pruna, D.~St\"ockinger and A.~Straessner,
  \emph{JHEP} {\bf 1209} (2012) 125
  [{\tt arXiv:1205.2692}].

\bibitem{Semenov:2010qt}
  A.~Semenov,
  [{\tt arXiv:1005.1909}].

\bibitem{Belyaev:2012qa}
  A.~Belyaev, N.~D.~Christensen and A.~Pukhov,
  \emph{Comput.\ Phys.\ Commun.\ } {\bf 184} (2013) 1729
  [{\tt arXiv:1207.6082}].

\bibitem{Pumplin:2002vw}
  J.~Pumplin, D.~R.~Stump, J.~Huston, H.~L.~Lai, P.~M.~Nadolsky and W.~K.~Tung,
  \emph{JHEP} {\bf 0207} (2002) 012
  [{\tt hep-ph/0201195}].

\bibitem{Alwall:2014hca}
  J.~Alwall, R.~Frederix, S.~Frixione, V.~Hirschi, F.~Maltoni, O.~Mattelaer, H.-S.~Shao and T.~Stelzer {\it et al.},
  \emph{JHEP} {\bf 1407} (2014) 079
  [{\tt arXiv:1405.0301}].

\bibitem{Conte:2012fm}
  E.~Conte, B.~Fuks and G.~Serret,
  \emph{Comput.\ Phys.\ Commun.\ } {\bf 184} (2013) 222
  [{\tt arXiv:1206.1599}].

\bibitem{Hermann:2012fc}
  T.~Hermann, M.~Misiak and M.~Steinhauser,
  \emph{JHEP} {\bf 1211} (2012) 036
  [{\tt arXiv:1208.2788}].

\bibitem{Chatrchyan:2012jua}
  S.~Chatrchyan {\it et al.}  [CMS Collaboration],
  \emph{JINST} {\bf 8} (2013) P04013
  [{\tt arXiv:1211.4462}].

\bibitem{Chatrchyan:2012zz}
  S.~Chatrchyan {\it et al.}  [CMS Collaboration],
  \emph{JINST} {\bf 7} (2012) P01001
  [{\tt arXiv:1109.6034}].

\bibitem{Bruning:2002yh}
  O.~S.~Bruning, R.~Cappi, R.~Garoby, O.~Grobner, W.~Herr, T.~Linnecar, R.~Ostojic and K.~Potter {\it et al.},
  CERN-LHC-PROJECT-REPORT-626.

\end{thebibliography}
\end{document}